\documentclass[12pt]{iopart}
\usepackage{epsfig}
\usepackage{dcolumn}
\usepackage{bm}
\usepackage{tikz}
\usepackage{graphicx, graphics, color}

\newcommand{\be}{\begin{equation}}
\newcommand{\ee}{\end{equation}}
\newcommand{\beq}{\begin{eqnarray}}
\newcommand{\eeq}{\end{eqnarray}}

\begin{document}

\title{Optimization  of a three-terminal non-linear heat nano-engine}
\author{Barbara Szukiewicz}
\ead{barbaraszukiewicz@gmail.com}
\address{Institute of Physics, 
Maria Curie-Sk{\l}odowska University, 
20-031 Lublin, Poland}
\author{Ulrich Eckern}
\ead{ulrich.eckern@physik.uni-augsburg.de}
\address{Institute of Physics, University of Augsburg, 86135 Augsburg, Germany}
\author{Karol I. Wysoki\'nski}
\ead{karol@tytan.umcs.lublin.pl}
\address{Institute of Physics, Maria Curie-Sk{\l}odowska University, 
20-031 Lublin, Poland,
and \\
Institute of Physics, University of Augsburg, 86135 Augsburg, Germany}

\date{\today}

\begin{abstract}
Charge and heat transport in nano-structures is dominated by non-equilibrium effects
which strongly influence their behaviour. These  effects are studied
in a setup consisting of three external leads, one of which is considered as a heat reservoir 
and is tunnel-coupled to two cold electrodes $via$ two independently controlled quantum dots.
The energy flow from the hot electrode together with energy filtering provided by quantum dots  
leads to a voltage bias between the cold electrodes. 
The heat and charge currents in the device effectively flow 
in mutually perpendicular directions, allowing for their independent control. 
The non-equilibrium screening changes the values of the system parameters 
needed for its optimal performance but  leaves the 
maximal output power and efficiency unchanged. Our results are important from the
theoretical point of view as well as for the practical implementation and the control of the 
proposed heat engine.
\end{abstract}
\maketitle
\section{Introduction}
The efficient harvesting of  waste heat is one of the most important challenges of modern
technology both at large and small scales. Combining the effective cooling of the
integrated circuits with simultaneous conversion of a part of the released heat into 
electric power could revolutionize electronics. 
On the road towards effective heat nano-engines a number of important findings 
have to be noticed. Among them, the observation of the importance of energy filtering~\cite{mahan1996},
and the independent control of heat and charge flow~\cite{sanchez2011,sothmann2012,jordan2013,mazza2014,mazza2015}
are  of great interest.  

\begin{figure}
\centerline{\includegraphics[width=0.50\linewidth]{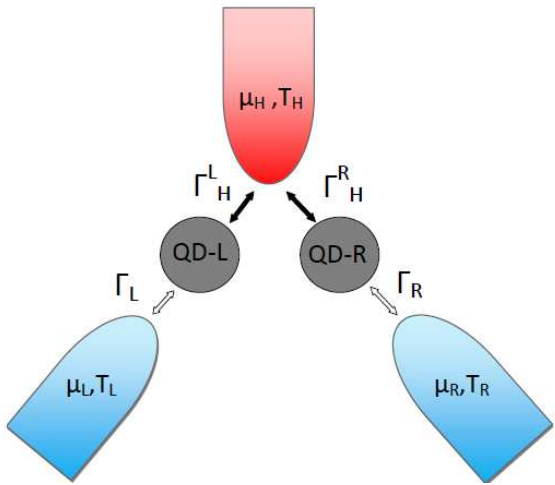}}
\caption{The structure of the heat engine. The left and right current junction contain
 quantum dots. The temperature of the hot reservoir equals $T_H$, while the two other reservoirs
are assumed to have the same temperature $T_L=T_R<T_H$. The chemical potentials of the electrodes
$\mu_{L,R,H}$ may be changed by the external voltage or as a result of the temperature difference
between hot and cold electrodes. }
\label{fig3the}
\end{figure}

These observations are at the heart of our approach as
we use quantum dots as efficient energy filters and the three-terminal setup for
independent control of heat and charge flow. The setup we are considering 
is shown in Fig.~\ref{fig3the}. It consists of two independently tuned quantum dots and three
external terminals. The left and right current junctions contain a 
quantum dot. The central terminal is the hot one. It can be considered as 
a cavity~\cite{jordan2013} connected to an external heat bath. The temperature of the hot 
reservoir equals $T_H$, while the two other reservoirs
are assumed to have lower temperatures $T_L$ and $T_R$, respectively. The chemical potentials of the electrodes
$\mu_{L,R,H}$ may be changed by the external voltage or as a result of the temperature difference
between hot and cold electrodes. 

Experimentally similar systems \cite{edwards1993} have been already produced and implemented
in electronic refrigerators~\cite{prance2009}, proved to be successful 
in cooling in the mK temperature range, and recently shown to be efficient 
heat harvesters \cite{hartmann2015,roche2015,thierschmann2015}. 
Thermoelectric nano-engines with quantum dots tunnel-coupled to external electrodes in two- and
three-terminal geometry have been proposed as effective heat to electricity 
converters~\cite{zippilli2009,rutten2009, esposito2009,humphrey,broeck,gaveau2010,entin-wohlman2010,jiang2012, ruokola2012,kennes2013,jiang2014,crepieuux2015,whitney2015,mintchev2015,yamamoto2015}. 
Note that besides quantum dots also molecules
\cite{sadeghi2015} and nano-wires \cite{bosisio2015} are 
 useful elements for efficient energy harvesting at the nano-scale.   
The field of thermoelectric energy harvesting with quantum dots has been been 
recently reviewed~\cite{sothmann2015,benenti2013}, while a more 
general discussion related to energy harvesting can be found in \cite{radousky2012}. 
 
Consider the system shown in Fig.~\ref{fig3the} with quantum dots energy levels differing by $\Delta E$.
Assume for a while that the tunnelling via quantum dots is possible only at sharp values of 
on-dot energies $\varepsilon_L$ and $\varepsilon_R$. In such a case an electron from the left lead
with energy $\varepsilon_L$ can tunnel into the H-lead, and an electron with energy  equal to $\varepsilon_R$
can tunnel from H to the right electrode. Thus each electron transferred between L and R electrode 
gains an energy $\Delta E= \varepsilon_R - \varepsilon_L$ from the H electrode. This process is possible 
if the temperature of the H electrode is the highest (hence hot electrode). The charge is transferred
between L and R electrodes at the cost of the heat from the H electrode. The dots act here as efficient energy 
filters, and charge effectively flows in direction perpendicular to the heat flow. In other words, 
the electron flow being a result of the temperature difference
between hot and cold electrodes gives rise to a voltage bias  between the two cold electrodes.
One can invert the perspective and argue that in the presence of the (not too large) voltage bias (load) 
between two cold electrodes the electron  flow at the cost of heat from  the hot electrode 
performs work against the bias.  
The value of the bias at which the charge currents stop to flow is called stopping bias,
and is denoted by $V_{\rm stop}$. The device operates as a heat engine in the voltage range $(0,V_{\rm stop})$.
The possibility of independent control of heat and charge flow is a main motivation
to consider the three terminal geometry.

A similar heat nano-engine has been recently optimised~\cite{jordan2013} for maximum power. 
The optimization involved the coupling strength 
of the dots to external electrodes, the ``energy gain'' $\Delta E$
and the voltage load between left and right electrodes. However, the authors~\cite{jordan2013} have not 
considered the non-equilibrium effects related to charge redistribution and screening, being of importance 
outside the linear transport approximation. 

Indeed, recent measurements of the thermoelectric voltage 
clearly show~\cite{svensson2013,svilans2015} that the observed non-linear effects are related 
to heating-induced renormalisation
of the dot energy levels. The other source of non-linearities, namely, the energy 
dependence of the transmission function has been found 
to play a negligible role. The renormalisations of the dot energy states by electric and thermal
gradients in a similar system have been recently studied  theoretically 
within the scattering approach~\cite{sanchez2013,meair2013,whitney2013}. 

In other words,  beyond the linear approximation the non-equilibrium screening
potentials start to play an importnat role. As a result, $e.g.$ 
the optimal set of parameters of an engine differs from that 
obtained in the theory which does not take such effects into account.
For finite voltage or temperature bias, the charges pile up in the electrodes
and quantum dots. Due to the long-range Coulomb interaction they screen other charges and
change the injection rates of particles from the electrodes~\cite{buttiker1993}.    
This observation is especially important for large temperature differences 
and large load voltages well beyond the validity of linear response.
 
At the nano-scale the issue of linear response is a tricky one. In principle it is
even not well defined. In bulk diffusive systems the small temperature difference 
between the far ends of the sample allows well defined local temperatures and an average temperature
gradient. Similarly a small bias usually leads to a  small gradient of the electrochemical  potential.
In nano-structures even small biases do not imply validity of linear response. To capture
non-linearity we shall use the non-equilibrium Green function approach to derive equations
for heat and charge currents and consider the effects of non-equilibrium screening of charges and their piling 
up in the electrodes~\cite{altshuler1985}. 

Working as an energy harvester the system converts the heat current $J$  
into power, $P = IV$, where $I$ is the charge current flowing between left and right (cold) electrodes. 
The voltage $V$ is used to power an external device (the load). The efficiency is defined
as the ratio between the useful power and the heat current flowing into the system, $\eta=P/J$.
The efficiency calculated in this way can be contrasted
with the ideal Carnot value $\eta_C=1-T_R/T_H$, and the Curzon-Ahlborn efficiency~\cite{curzon1974}
expected at maximum power, $\eta_{CA}=1-\sqrt{T_R/T_H}=1-\sqrt{1-\eta_C}\approx \eta_C/2+ \cdots$. 
 
The organization of the paper is as follows. In the next section we present the microscopic
model of the system at hand, and calculate the charge and heat currents using the non-equilibrium Green
function technique. The non-linear effects in transport are discussed in Sec.\ III. The results
of the optimization of the engine working well outside the linear regime are presented
in Sec.\ IV. We end with summary and conclusions (Sec.\ V). 

\section{Model and approach}
The Hamiltonian of the system is written as
\be
\hat{H}=\sum_{\lambda,\vec{k},\sigma}^{L,R,H}\varepsilon_{\lambda \vec{k} \sigma}n_{\lambda \vec{k} \sigma} + \sum _{i\sigma} (\varepsilon _{i}-e U_{i})n_{i\sigma} 
+\sum_{i\lambda \vec{k} \sigma} (\tilde{V}_{i\lambda \vec{k}} c^{\dagger}_{\lambda k\sigma} d_{i \sigma} 
+ \tilde {V}_{i\lambda \vec{k}}^* d^{\dagger}_{i\sigma} c_{\lambda \vec{k} \sigma}),
\label{ham1}
\ee
where $n_{\lambda \vec{k} \sigma}=c^{\dagger}_{\lambda \vec{k} \sigma}c_{\lambda \vec{k} \sigma}$~
and $n_{i\sigma}=d^{\dagger}_{i\sigma} d_{i \sigma}$
denote particle number  operators for the leads and the dots, respectively. 
The operators $c^{\dagger}_{\lambda k\sigma} (d^{\dagger}_{i\sigma})$ create electrons in
respective states $\lambda \vec{k}\sigma$ $(i\sigma)$ in the leads (on the dots).
Symbols $i=1, 2$ refer to the left and right dot, and 
$\lambda=L, R, H$ denote the left, right, and hot electrode, respectively. 
The dot energy levels $\varepsilon _{L}\equiv \varepsilon_1$ and $\varepsilon _R\equiv\varepsilon_2$ can
be easily tuned by the gate voltages. They are renormalised by the potentials $U_i$ which account for the
electron-electron repulsion. This interaction is considered here at the mean-field level.

The bare tunneling amplitudes between dot $i$ and electrode $\lambda$ are denoted by $V_{i\lambda\vec{k}}$. 
We introduce the symbol
\be
\tilde{V}_{i\lambda\vec{k}}=V_{i\lambda\vec{k}}[(1-\delta_{\lambda H})\delta_{\lambda i}+\delta_{\lambda H}],
\ee
which takes into account that left and right  leads
are coupled, respectively to the first $1$ ($=L$) and the second $2$ ($=R$) 
quantum dot, while both dots are coupled to the $H$ electrode. 

The current in the electrode $\lambda$ is calculated as time derivative of the average charge in that electrode 
$N_{\lambda}=\sum_{k\sigma} n_{\lambda\vec{k}\sigma}$:

\be
I_{\lambda}=-e\left\langle\frac{dN_{\lambda}}{dt}\right\rangle=-e\frac{d}{dt}\left\langle\sum_{\vec{k}\sigma}n_{\lambda \vec{k} \sigma}\right\rangle,
\label{current1}
\ee
where the symbol $\langle...\rangle$ denotes the statistical average.
Calculation of the heat flux follows that of the charge. From thermodynamics we 
know the relation between the (internal) energy $E$, heat $Q$, and work. Assuming that the
only work is related to the flow of mass $\mu dN$, one writes 
\be
\delta \dot{Q}= d\dot{E} -\mu d\dot{N}.
\ee
The dot over symbols denotes time derivative. 
Applying this equation to one of the electrodes, say $\lambda$, allows
to write the heat flux as
\be
J_\lambda=\frac{i}{\hbar}\langle[H_\lambda,\hat{H}]\rangle -\mu_\lambda\frac{i}{\hbar}\langle[N_\lambda,\hat{H}]\rangle,
\label{hcurrent}
\ee 
where
$H_\lambda=\sum_{\vec{k},\sigma}\varepsilon_{\lambda \vec{k} \sigma}n_{\lambda \vec{k} \sigma}$ 
is the  energy operator for the electrode $\lambda$. 
Evaluating the commutators and defining Keldysh ``lesser'' functions: 
\be
G^{<}_{i\sigma,\lambda\vec{k}\sigma} (t, t{'})\equiv i\langle c^{\dagger}_{\lambda\vec{k}\sigma} (t{'})d_{i\sigma} (t)\rangle,
\ee
\be
G^{<}_{\lambda\vec{k}\sigma,i\sigma} (t, t{'}) \equiv i\langle d^{\dagger}_{i\sigma} (t') c_{\lambda\vec{k}\sigma}(t)\rangle,
\ee
one gets
\beq
I_{\lambda}(t)=\frac{2e}{\hbar}\sum_{i\vec{k}\sigma}{\rm Re}\bigg[\tilde{V}_{i\lambda\vec{k}}G^{<}_{i\sigma,\lambda\vec{k}\sigma}(t,t)\bigg],
\label{curr1} \\
J_{\lambda}(t)=\frac{2e}{\hbar}\sum_{i\vec{k}\sigma}(\varepsilon_{\lambda \vec{k}}-\mu_\lambda) {\rm Re}
\bigg[\tilde{V}_{i\lambda\vec{k}}G^{<}_{i\sigma,\lambda\vec{k}\sigma}(t,t)\bigg].
\label{hcurr1}
\eeq
The final expressions for the stationary currents can easily be written in the general form~\cite{haug-jauho1996}
\beq
I_{\lambda}=2\frac{ie}{\hbar} \int\frac{dE}{2\pi} \sum_{ij} \Gamma^{\lambda}_{ij} (E)
\{ G^{<}_{ji} (E) + f_{\lambda}(E) [G^{r}_{ji}(E)-G^{a}_{ji}(E)]\}, \\
J_{\lambda}=2\frac{ie}{\hbar} \int\frac{dE}{2\pi} \sum_{ij} \Gamma^{\lambda}_{ij} (E)
(E-\mu_\lambda)\{ G^{<}_{ji} (E) + f_{\lambda}(E) [G^{r}_{ji}(E)-G^{a}_{ji}(E)]\},
\label{heat-curr}
\eeq
where 
\be
\Gamma^{\lambda}_{ij}(E)=2\pi\sum_{\vec{k}}\tilde{V}_{i\lambda \vec{k}}\tilde{V}^*_{j\lambda \vec{k}} \, \delta(E
-\varepsilon_{\lambda\vec{k}})
\ee
denotes the matrix of effective couplings of dots $(i,j)$ to the lead $\lambda$.
The factor 2 in the formulas for currents stems from the summation over spins. 
The heat current (\ref{heat-curr})  can be written as a difference between
the energy current $J^E_\lambda$ and the charge current $I_\lambda$:
\be
J_\lambda=J^E_\lambda-\mu_\lambda I_\lambda.
\ee

To calculate lesser Green function~\cite{haug-jauho1996} we use
the equation of motion method~\cite{niu1999}. From now on, we shall work in units with Planck constant 
$\hbar=1$ and Boltzmann constant $k_B=1$.
It is convenient to define the frequency dependent dot matrix Green function $\hat{G}^<(\omega)$ with elements
\be
G^<_{ij}(\omega)=\langle\langle d_{i\sigma}|d^\dagger_{j\sigma}\rangle\rangle^{<}_{\omega}.
\label{matrixlGF}
\ee
For the definition of $\langle\langle ...|...\rangle\rangle$, see Ref.~\cite{niu1999}. 
After some algebra one finds
\be
\hat{G}^{<} (\omega)=\hat{G}^{r}(\omega)\hat{\Sigma}^{<}(\omega)\hat{G}^{a}(\omega),
\label{lesserG}
\ee
where the matrix lesser self-energy is given by
\be
\hat{\Sigma}^{<}(\omega)=\sum_{\lambda\vec{k}} g^{<}_{\lambda k} (\omega)\left(\begin{array}{cc}|\tilde{V}_{1\lambda k}|^{2} , \tilde{V}^*_{1\lambda k}\tilde{V}_{2\lambda k}\\\tilde{V}^*_{2\lambda k}\tilde{V}_{1\lambda k}, |\tilde{V}_{2\lambda k}|^{2}\end{array}\right).
\ee
The equation for the retarded matrix Green function~\cite{zubariev1960} $G^{r}_{ij}(\omega)$
can be written in explicit form as
\beq
\left( \begin{array}{cc}\omega-\varepsilon_{1}+eU_{1}-\Sigma^r_{11};-\Sigma^r_{12}\\
-\Sigma^r_{21};\omega-\varepsilon_{2}+eU_{2}-\Sigma^r_{22}\end{array}\right)
\left(\begin{array}{cc}G^r_{11}(\omega),G^r_{12}(\omega)\\G^r_{21}(\omega),G^r_{22}(\omega)\end{array}\right) 
=\left( \begin{array}{cc}1,0\\0,1\end{array}\right),
\label{GFM}
\eeq
with the retarded self-energy
\be
\Sigma^{r}_{ij} (\omega)=\sum_{\lambda\vec{k}}\frac{\tilde{V}^*_{i\lambda\vec{k}}\tilde{V}_{j\lambda\vec{k}}}{\omega-\varepsilon_{\lambda\vec{k}}+i0}.
\ee
The matrix inversion gives all the components of the required retarded function. 
In the wide band limit approximation one replaces the retarded self-energy by its imaginary part only:
\be
\Sigma^{r}_{ij} (\omega)\approx -i\pi \sum_{\lambda\vec{k}}\tilde{V}^*_{i\lambda k}\tilde{V}_{j\lambda k} \, \delta (\omega-\varepsilon_{\lambda k})
=-\frac{i}{2}\sum_\lambda\Gamma^\lambda_{ij} (\omega),
\ee
and neglects the frequency dependence of $\Gamma^\lambda_{ij} (\omega)=\Gamma^\lambda_{ij}$.

\section{Non-linear effects}
The long-range nature of the Coulomb interactions is responsible for the back-reaction 
of the non-equilibrium  charge
distribution onto the transport properties of the device. In the Hamiltonian (\ref{ham1}) this is represented
by the screening potentials $U_i$. Their values depend on the thermoelectric configuration, $i.e.$ voltages and
temperatures of all electrodes.  This effect has been considered  in mesoscopic normal 
systems first by Altshuler and Khmelnitskii~\cite{altshuler1985}, and later by 
B\"{u}ttiker and coworkers~\cite{buttiker1993,buttiker1997}, 
and others~\cite{ma1998,sheng1998,hernandez2009,sanchez2013,hershfield2013,whitney2013}. 
It has been also explored in metal-superconductor 
two-terminal~\cite{wang2001,hwang2015} and three-terminal junctions~\cite{michalek2015}. 

Here we follow Ref.~\cite{wang2001} and others~\cite{hwang2015,michalek2015}, assuming that 
the long-range interactions modify the on-dot 
energies $\varepsilon_i$, changing them into $\varepsilon_i - e U_i$. In equilibrium the potentials $U_i$ 
have constant values (independent of $V_\lambda$ and $\Delta T_\lambda)$,
which we denote by $U_{i,{\rm eq}}$. 
In the presence of applied voltages $V_\lambda$ and temperature biases $\Delta T_\lambda$, 
the deviations $\delta U_i = U_i - U_{i,{\rm eq}}$, in lowest order, are linear functions~\cite{wang2001}
of $V_\lambda$ and $\Delta T_\lambda$.

Thus we write for the potential on each dot $i=1,2$:
\begin{equation} 
\delta U_i=\sum_{\lambda}^{L,R,H}\left[\left(\frac{\partial U_i}{\partial V_{\lambda}}\right)_{0}V_\lambda+
\left(\frac{\partial U_i}{\partial \Delta T_{\lambda}}\right)_{0}\Delta T_{\lambda}\right]  
+ ...~,
\label{dudvdt}
\end{equation}
where the subscript zero indicates that the partial derivatives have to be evaluated with 
all $V_\lambda, \Delta T_\lambda$ set to zero,
and the dots denote higher order terms.
The charge densities on the dots $\langle n_i\rangle$ also depend on the temperature and voltage 
bias as well as on the potentials $U_i$.
Expanding  to lowest order in these parameters we get
\begin{eqnarray} 
\delta \langle n_i\rangle&=&\langle n_i\rangle-\langle n\rangle_{i,{\rm eq}} \nonumber \\
&=&\sum_\lambda^{L,R,H}\left[\left(\frac{\partial \langle n_i\rangle}{\partial V_\lambda}\right)_0 V_\lambda +
\left(\frac{\partial \langle n_i\rangle}{\partial \Delta T_\lambda}\right)_0\Delta T_\lambda\right]    
-\sum_j \Pi_{ij} \delta U_j + \ldots . 
\label{dndU}
\end{eqnarray}
The above equation defines the Lindhard matrix function as
\begin{equation}
\Pi_{ij}=-\left(\frac{\partial\delta\langle n_i\rangle}{\partial\delta U_j}\right)_0.
\end{equation}
The derivatives can be easily calculated by noting that
\be
\langle n_i\rangle=\sum_\sigma\langle d_{i\sigma}^\dagger d_{i\sigma}\rangle=\frac{-i}{\pi}\int dE G^<_{ii}(E),
\ee
and using equation (\ref{lesserG}).   
Another relation between charges and potentials defines the capacitance  matrix $\hat{C}$ of the system: 
\begin{equation}
\delta \langle n_i\rangle= \sum_j C_{ij}\delta U_j.
\label{cap}
\end{equation}
The equations  (\ref{dndU}) and (\ref{cap}) are easily solved,
and one finds explicit expressions for the characteristic potentials 
\begin{eqnarray}
&&u_{i,\lambda} \equiv \left(\frac{\partial U_i}{\partial V_{\lambda}}\right)_{0}=
\sum_j [(\hat{C}+\hat{\Pi})^{-1}]_{ij}\left(\frac{\partial \langle n_j\rangle}{\partial V_\lambda}\right)_0, \\
&&z_{i,\lambda}\equiv \left(\frac{\partial U_i}{\partial \Delta T_{\lambda}}\right)_{0}=
\sum_j [(\hat{C}+\hat{\Pi})^{-1}]_{ij}\left(\frac{\partial \langle n_j\rangle}{\partial \Delta T_{\lambda}}\right)_0. 
\label{uz-pot}
\end{eqnarray}
The knowledge of the characteristic potentials allows to calculate how the temperature difference 
between hot and cold electrodes and voltages modify the potentials $U_i$ of
the dots. These changes, in turn, affect the heat and charge currents flowing in the system.
For the explicit results presented below, we will assume the small capacitance limit, $\hat{C}\approx 0$.

Due to the approximation in Eq. (\ref{dudvdt}) our approach is called ``weakly non-linear'', 
as we do not consider higher order corrections.

\section{Results}

We assume that the left and right electrodes of the system (cf.\ Fig.~\ref{fig3the})
have the same temperature, $T_L = T_R$. The temperature of the hot electrode, which is kept fixed from now on, is denoted by $T_H$
($i.e.$ $\Delta T_H = 0$). In addition, $\Delta T_R = \Delta T_L = T_H-T_R \equiv \Delta T$, and the average temperature of the system
is $T=(T_H+T_R)/2$. The current does not flow into or out of the hot electrode which is grounded ($I_H=0$). This means that
charge conservation written in the form $I_L+I_R=0$ serves as a condition for the chemical potential $\mu_H$
of the hot electrode. The energy conservation $J^E_H+J^E_L+J^E_R=0$ may serve as a condition
for the actual temperature of that electrode. We shall take another
point of view and assume that the hot electrode serves as an energy reservoir  
characterised by the constant temperature $T_H$. The heat current $J_H$ flows out 
of it towards the $L$ and $R$ electrodes. For $\Delta E=\varepsilon_R-\varepsilon_L \ne 0$ the electrons entering the left 
electrode at energy $\varepsilon_L$ leave the right one at energy $\varepsilon_R$. As a result the voltage
$V$ appears between both electrodes.

To facilitate the calculations we impose additional conditions. The couplings of the quantum dots to external leads fulfil
$\Gamma^L_{ij}=\Gamma_L\delta_{i1}\delta_{j1}$, $\Gamma^R_{ij}=\Gamma_R\delta_{i2}\delta_{j2}$,
and we assume the matrix $\Gamma^H_{ij}$  to be symmetric with elements $\Gamma^H_{11}$, $\Gamma^H_{12}=\Gamma^H_{21}$
and $\Gamma^H_{22}$.  We tune the positions of 
the dots' energy levels symmetrically with respect to the chemical potential $\mu_H$.
Also the voltages are assumed to be symmetrical with respect to $\mu_H$, $i.e.$
$\mu_L=\mu_H-eV/2$ and $\mu_R=\mu_H+eV/2$. Such choice of parameters assures that $I_H=0$, and $I=I_L=-I_R$. 
For an arbitrary set of parameters not fulfilling the above symmetries, $\mu_H$ has to be calculated from the
condition $I_H=0$.

\subsection{Linear transport parameters, power factor, and efficiency\label{subs:lin}}
From charge  and heat currrents we calculate linear  transport characteristics of the device 
including charge ($G$) and thermal ($\kappa$) conductances, and Seebeck coefficient $S$.
For convenience we change the notation and denote the charge current as $I_1$ and the heat current as $I_2$. 
Expanding the currents 
to linear order in  bias and temperature forces  $X_1=eV/T$ and  $X_2=\Delta T/T^2$, we write 
the fluxes in standard notation:
\beq
I_1=L_{11}X_1+L_{12}X_2, \\ 
I_2=L_{21}X_1+L_{22}X_2.
\eeq 
 The transport coefficients are given by the parameters $L_{ij}$. In accordance with
standard definitions one finds
the conductance $G=(I_1/V)_{\Delta T=0}=L_{11}/T$, the Seebeck coefficient which we ocasionally refer to as thermopower 
$S=-(V/\Delta T)_{I_1=0}=L_{12}/(TL_{11})$,  
and the thermal conductance $\kappa=(I_2/\Delta T)_{I_1=0}=(L_{22}-L_{12}L_{21}/L_{11})/T^2$.
The Peltier coefficent  defined as $\Pi=(I_2/I_1)_{\Delta T=0}$ is given by $L_{21}/L_{11}$.

\begin{figure}
\centerline{\includegraphics[width=0.45\linewidth]{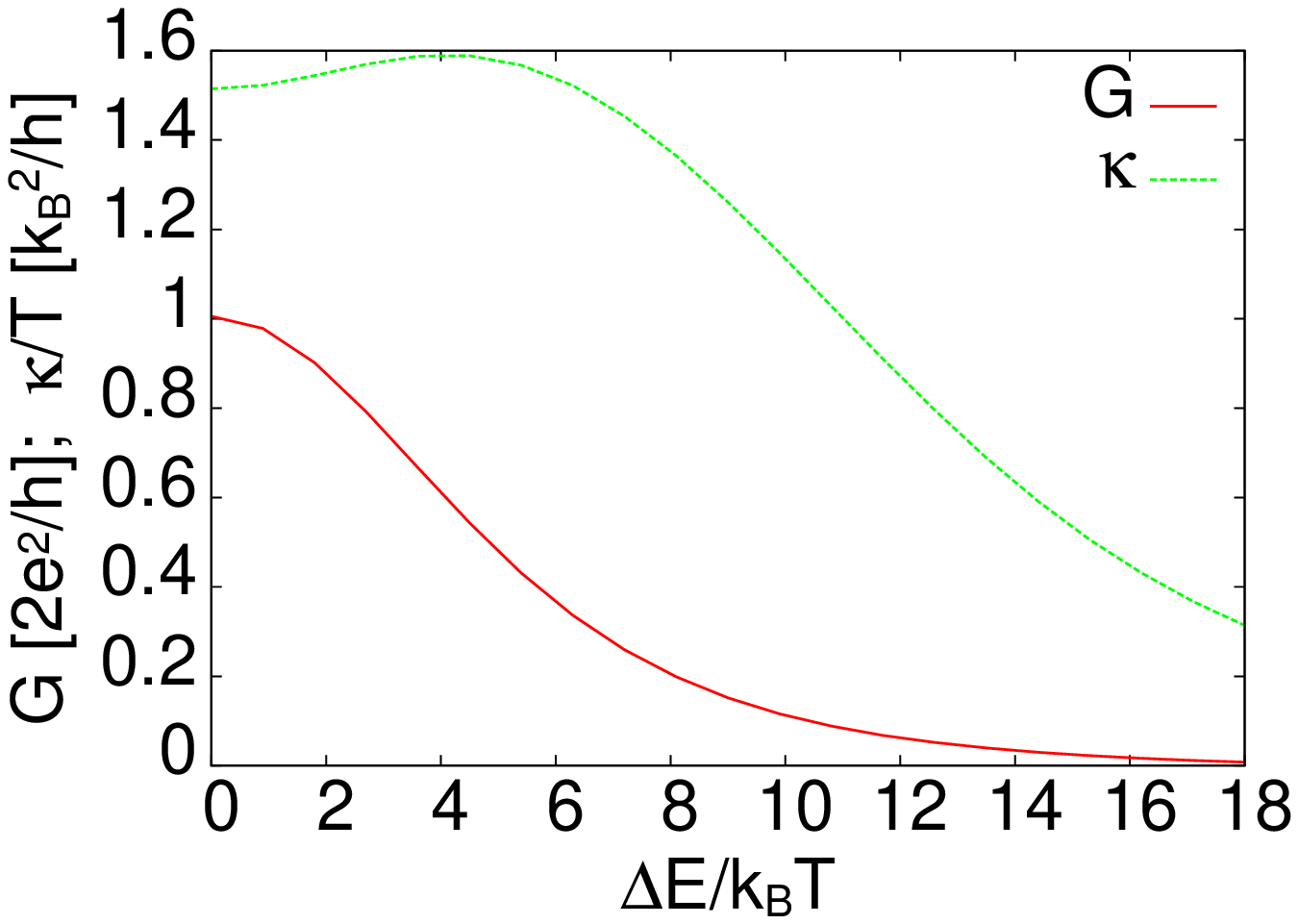}
\includegraphics[width=0.45\linewidth]{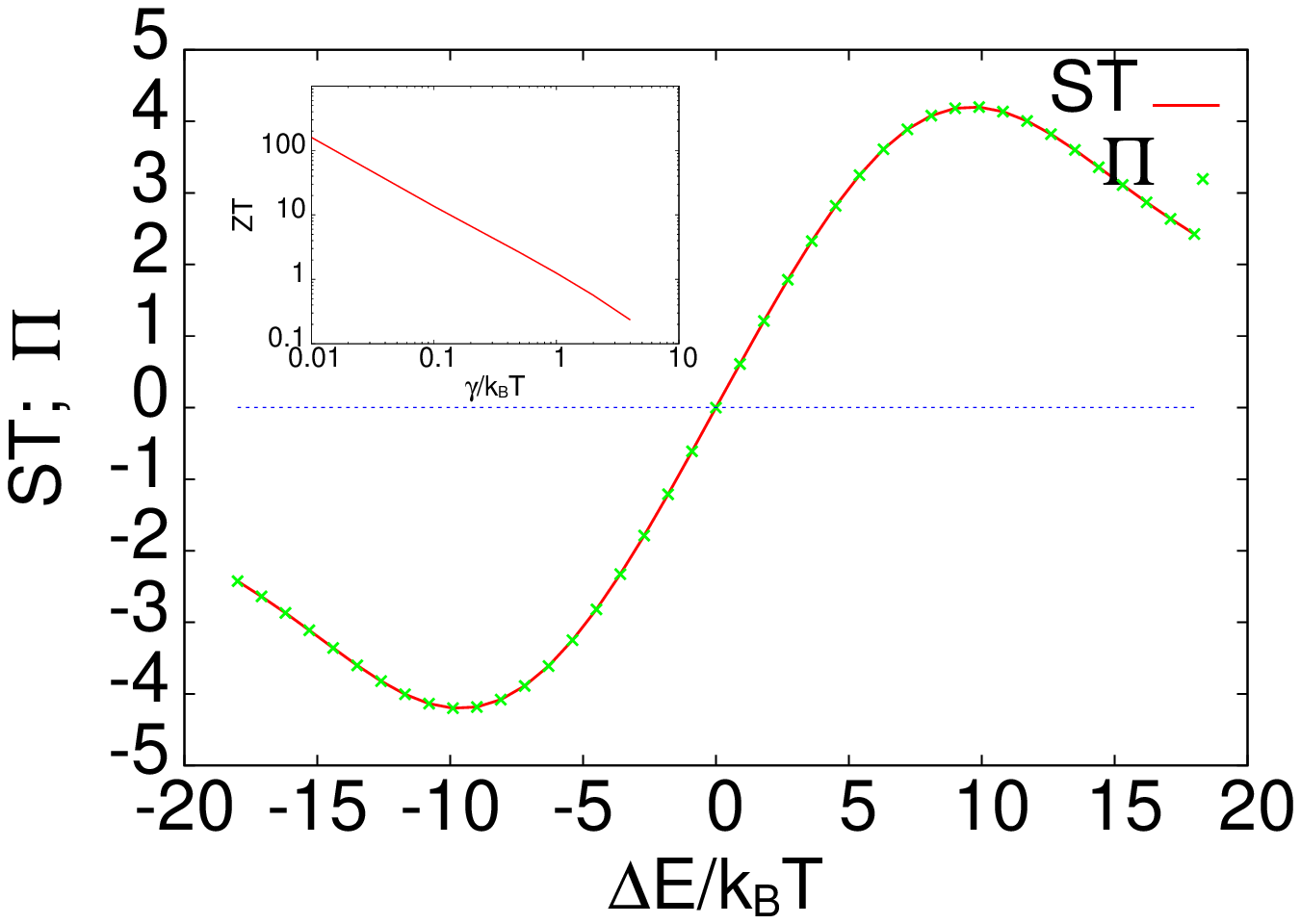}}
\caption{Linear conductance $G$ and thermal conductance $\kappa/T$ are shown
(left panel) as a function of the energy difference $\Delta E/k_BT$ for $\gamma=k_BT$.
The right panel shows the thermopwer $S$ multiplied by the temperature $T$, and the Peltier coefficient. 
Clearly $ST=\Pi$, in agreement with the Onsager reciprocity relations. The inset shows the 
$\gamma$ dependence of the thermoelectric figure of merit $ZT=GS^2T/\kappa$. }
\label{fig-lin}
\end{figure}

The combination of these parameters defines the thermoelectric
figure of merit $ZT=GS^2T/\kappa$, which enters the expression for the efficiency $\eta_{\rm lin}$
of the  thermoelectric heat engine~\cite{benenti2013}:
\be
\eta_{\rm lin}=\eta_C\frac{\sqrt{ZT+1}-1}{\sqrt{ZT+1}+1}.
\label{eff-lin}
\ee
The expected efficiency based on linear coefficients will serve as a reference below.
In  Fig.~\ref{fig-lin} we show the dependence of the linear conductance $G$, thermal conductance $\kappa$,  
Seebeck $S$ and Peltier $\Pi$ coefficients as well as power factor $GS^2$ 
$vs.$~$\Delta E/k_BT$ for the system with all couplings equal, $\Gamma_L=\Gamma_R=\Gamma^H_{ij}=\gamma=k_BT$, $i,j=1,2$.
As we shall see later this value of the coupling ($\gamma/k_BT=1$) leads to the maximum power.
The conductances $G$ and $\kappa$ are shown for positive values of $\Delta E$, as they are 
even functions of this parameter. On the other hand, both the Seebeck and Peltier effects are sensitive
probes of the electron or hole dominated transport, so they change sign as functions of $\Delta E$.
In the figure $S$ has been plotted, which by the Onsager relation equals $\Pi$.

The linear transport coefficients and the thermoelectric figure of merit $ZT$ depend on 
$\Delta E$ and $\gamma$. While both, the conductance $G$ and thermal conductance $\kappa$ 
monotonically increase, the thermopower decreases with increasing coupling $\gamma$. 
The increase of $G$ and $\kappa$ with $\gamma$ is related to the fact that the currents (\ref{curr1}),
 (\ref{hcurr1}) are proportional to $\gamma$.
To understand the decrease of the thermopower with $\gamma$ it is useful to recall
that in nano-structures $S$ is directly related to the slope of the density of states
at the Fermi energy~\cite{scheibner2005}, which is higher for smaller $\gamma$.  
In the linear approach the thermoelectric figure of merit yields the 
efficiency of the engine. Taking into account the behaviour of $ZT$ on $\gamma$
shown in the inset in the right panel of Fig.~\ref{fig-lin} 
and the formula (\ref{eff-lin}),  we conclude that maximum efficiency, approaching 
 the Carnot value $\eta_C$, is obtained for vanishingly small $\gamma$. However, the power of the
engine approaches zero value, rendering the engine practically useless.

\begin{figure}
\centerline{\includegraphics[width=0.50\linewidth]{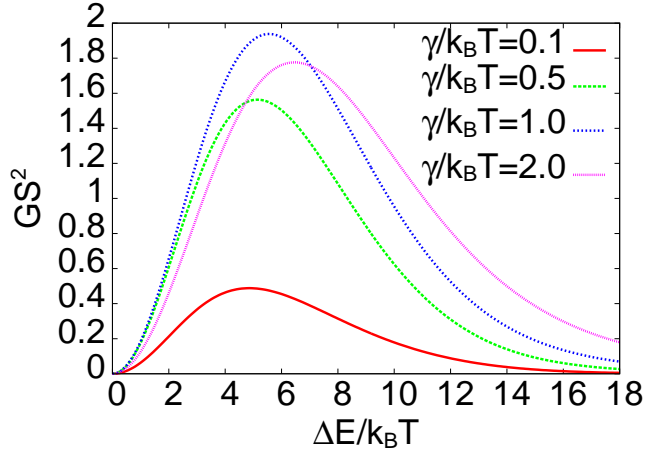}}
\caption{The dependence
of the power factor $GS^2$ on  $\Delta E$ for a number of $\gamma$ values. 
Note the non-monotonous dependence of its maximum value on $\gamma$ accompanied by 
the change of $\Delta E$ for which maximum is achieved.}
\label{power-fac-gam}
\end{figure}

We are interested in achieving the maximum power, which can be realised by optimising
$\Delta E$ and $\gamma$. In the linear theory the appropriate 
parameter characterising the obtained power is the power factor defined as $GS^2$. 
Its dependence on $\Delta E$ for a few values of $\gamma$ is shown in Fig.~\ref{power-fac-gam}.   
As we shall see in the next section the power factor shows a dependence
on $\Delta E$ and $\gamma$ qualitatively similar to that found in the 
exact non-linear approach, however, with a few important differences to be discussed later.

\subsection{Non-linear transport: maximum power and efficiency of the heat nano-engine\label{subs:nonlin}}

\begin{figure}
\centerline{\includegraphics[width=0.450\linewidth]{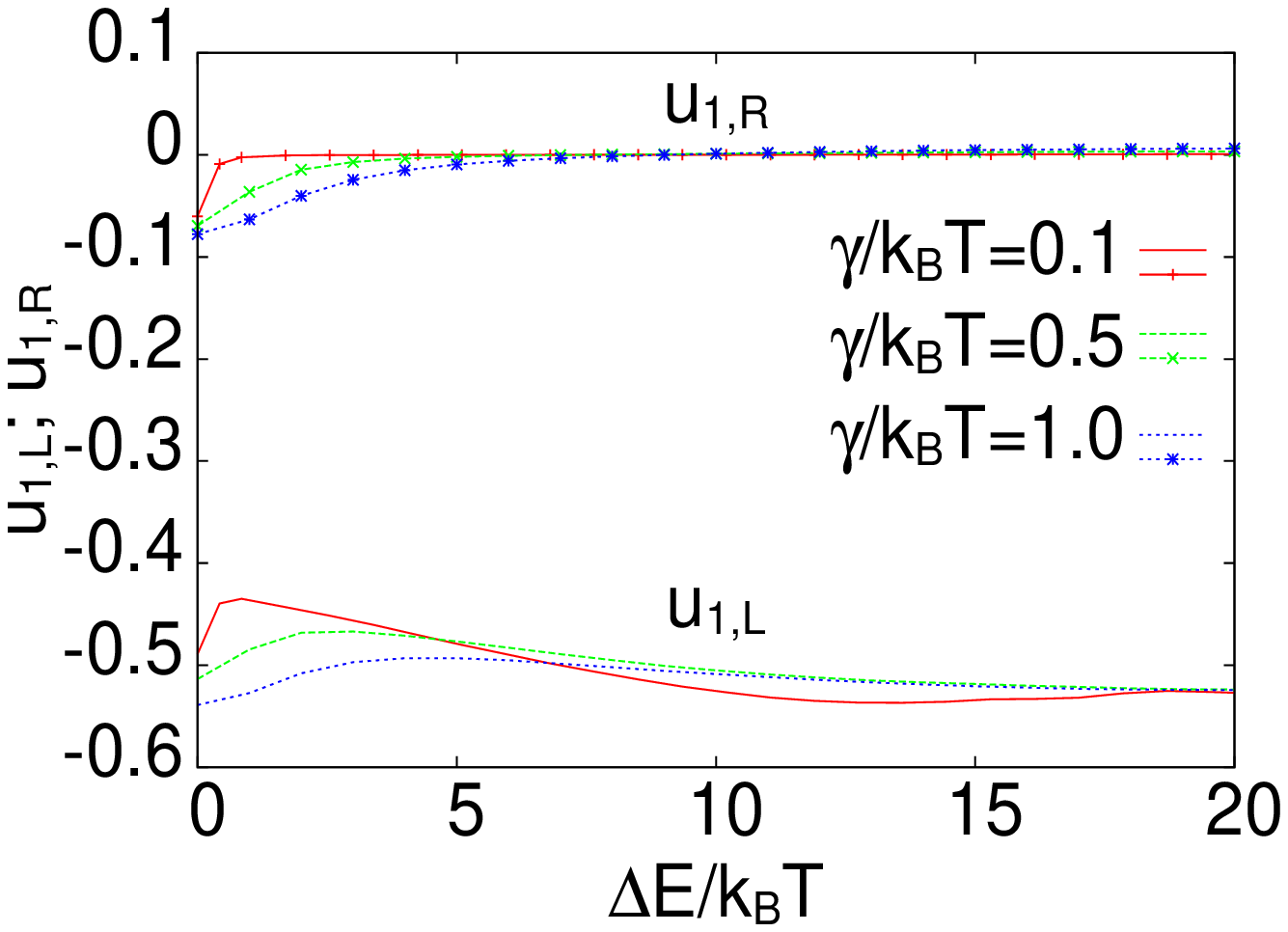}
\includegraphics[width=0.450\linewidth]{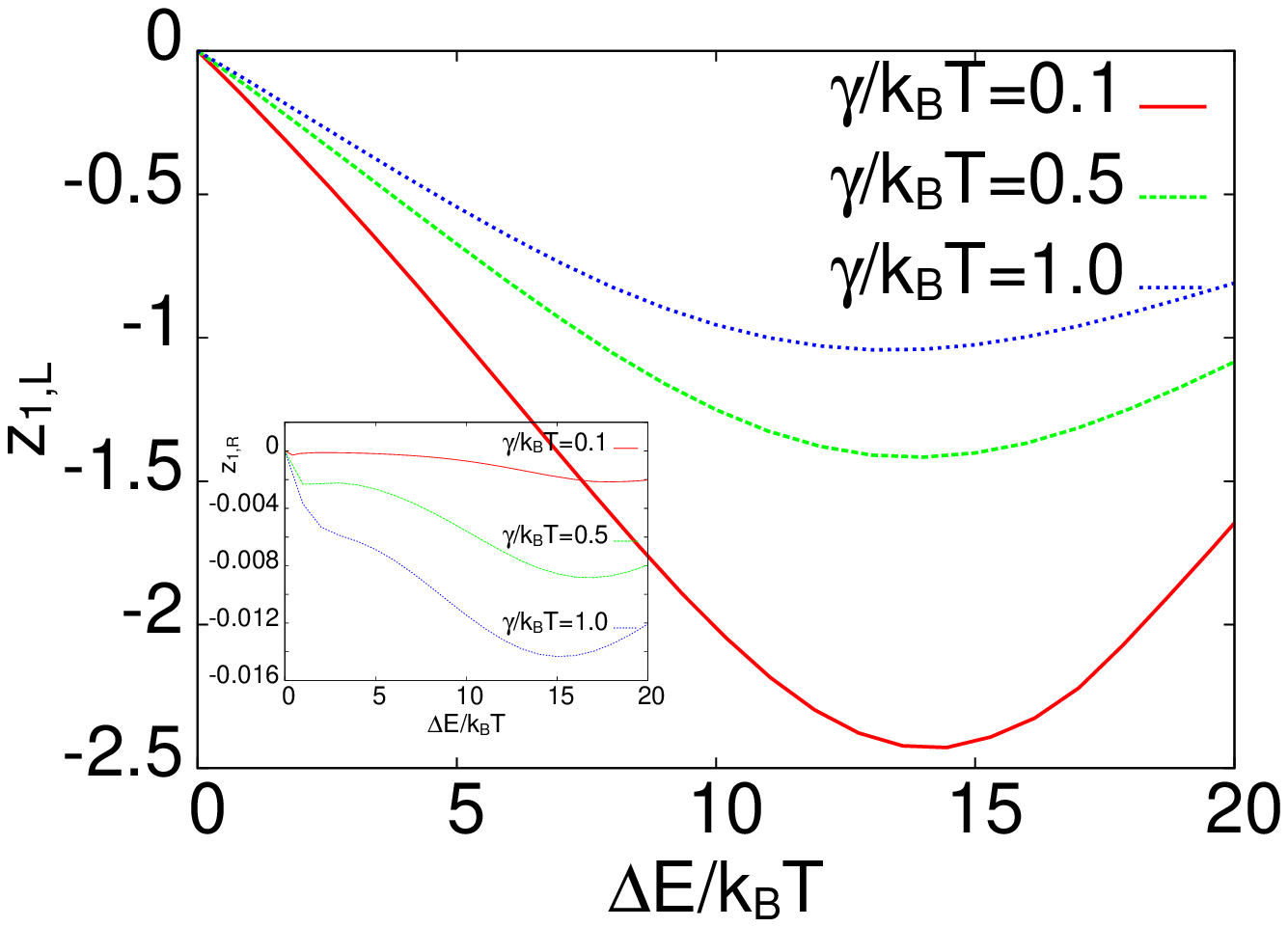}}
\caption{The dependence
of the characteristic potentials $u_{1,L}$, $u_{2,L}$ (left panel) and $z_{1,L}$ (right panel) 
on the  energy level difference $\Delta E=\varepsilon_R-\varepsilon_L$ for a number of $\gamma$ values. 
The inset in the right panel shows $z_{1,R}$. 
The energies are measured in units of $k_BT$, and the calculations have been done for $\Delta T/T=0.3$.}
\label{fig-uiL-ziLH12=0}
\end{figure}

We start the presentation of the results obtained with non-equilibrium screening
potentials taken into account by showing the parameters
$u_{i,\lambda}$ and $z_{i,\lambda}$ and their dependence on the 
couplings $\Gamma_{ij}$ and the energy difference $\Delta E=\varepsilon_R-\varepsilon_L$. These together
with the voltage load $V$ are the main optimization parameters.

In Fig.~\ref{fig-uiL-ziLH12=0} we show the dependence 
of the characteristic potentials $u_{1,L}$, $u_{2,L}$ and $z_{1,L}$, $z_{1,R}$ on  $\Delta E$. 
The behaviour of the other
parameters $u_{1,R}$, $u_{2,R}$, $z_{2,L}$ and $z_{2,R}$ can be inferred from their symmetries. 
For the symmetric system we are 
dealing with, the parameters $u_{i,\lambda}$  are even functions of $\Delta E$ and fulfil the relations: 
$u_{1,L}(\Delta E)=u_{2,R}(\Delta E)$, $u_{1,R}(\Delta E)=u_{2,L}(\Delta E)$.
The $z$ parameters 
are antisymmetric functions of $\Delta E$, $e.g.$ $z_{1,L}(\Delta E)=-z_{1,L}(-\Delta E)$
and are related as: $z_{1,L}(\Delta E)=-z_{2,R}(\Delta E)$, and $z_{1,R}(\Delta E)=-z_{2,L}(\Delta E)$.  
The ``off-diagonal'' characteristic potentials are much smaller than the ``diagonal''  
ones. This is especially true for the thermal potentials, $z_{1,R}$ and $z_{2,L}$, which are two orders 
of magnitude smaller than $z_{1,L}$ and $z_{2,R}$. 

The symbol $\gamma$ used in the figure denotes the 
common value of the coupling parameters, $\Gamma_L=\Gamma_R=\Gamma^H_{11}=\Gamma^H_{22}=\Gamma^H_{12}=\Gamma^H_{21}=\gamma$.
The  calculations have been performed for $\Delta T/T=0.3$.  
The diagonal characteristic potentials show a stronger dependence 
on $\Delta E$ and larger variation  for smaller values of the couplings $\gamma$. 
The amplitude strongly increases with increasing $\Delta T/T$.

If the couplings $\Gamma^H_{12},~\Gamma^H_{21}$ are assumed to vanish, then also 
the off-diagonal characteristic potentials $u_{1,R}$, $u_{2,L}$, and $z_{1,R}$, $z_{2,L}$ vanish. 
This fact, however, has only a small effect on the performance of the engine.
 
\begin{figure}
\centerline{\includegraphics[width=0.450\linewidth]{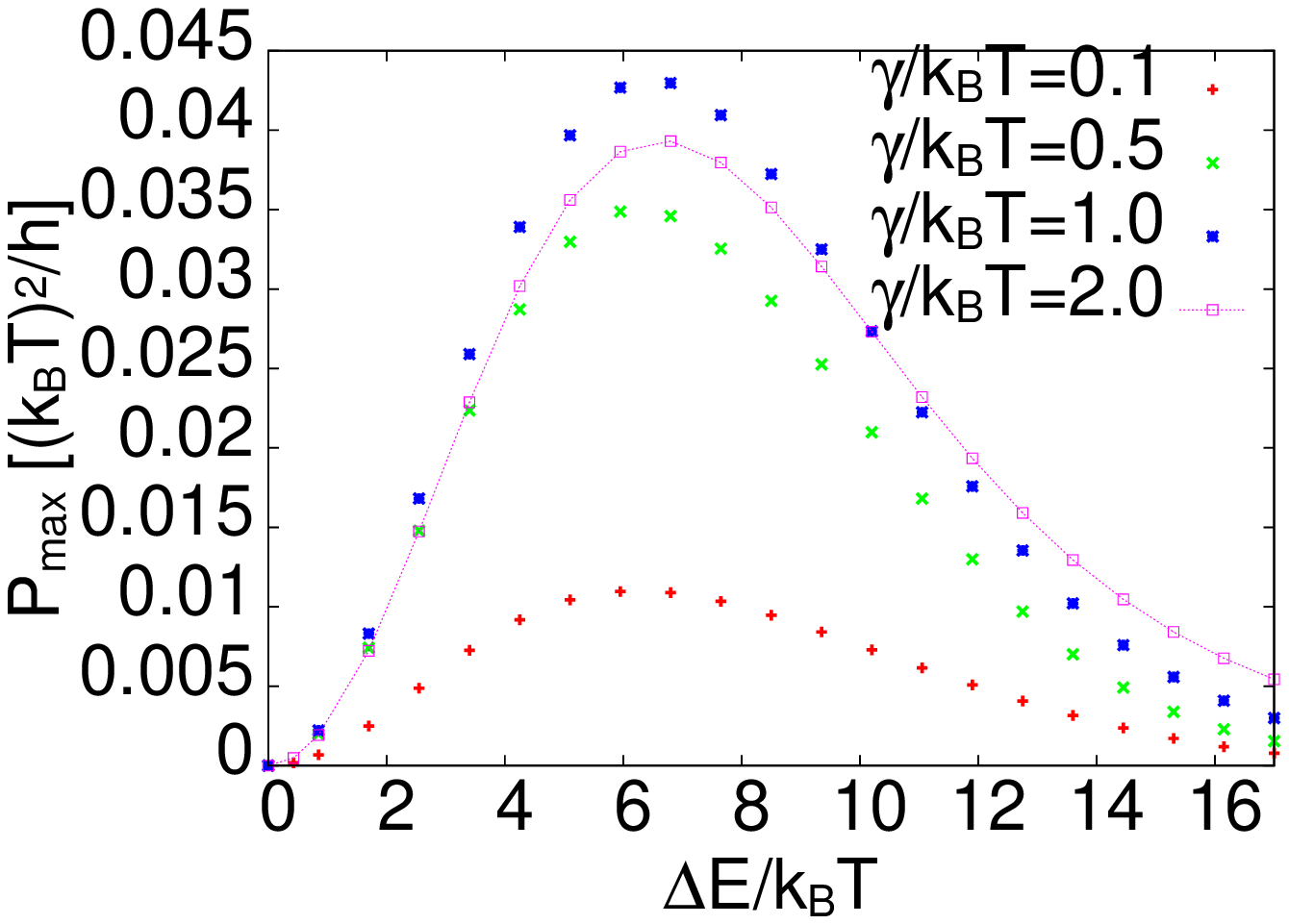}
\includegraphics[width=0.450\linewidth]{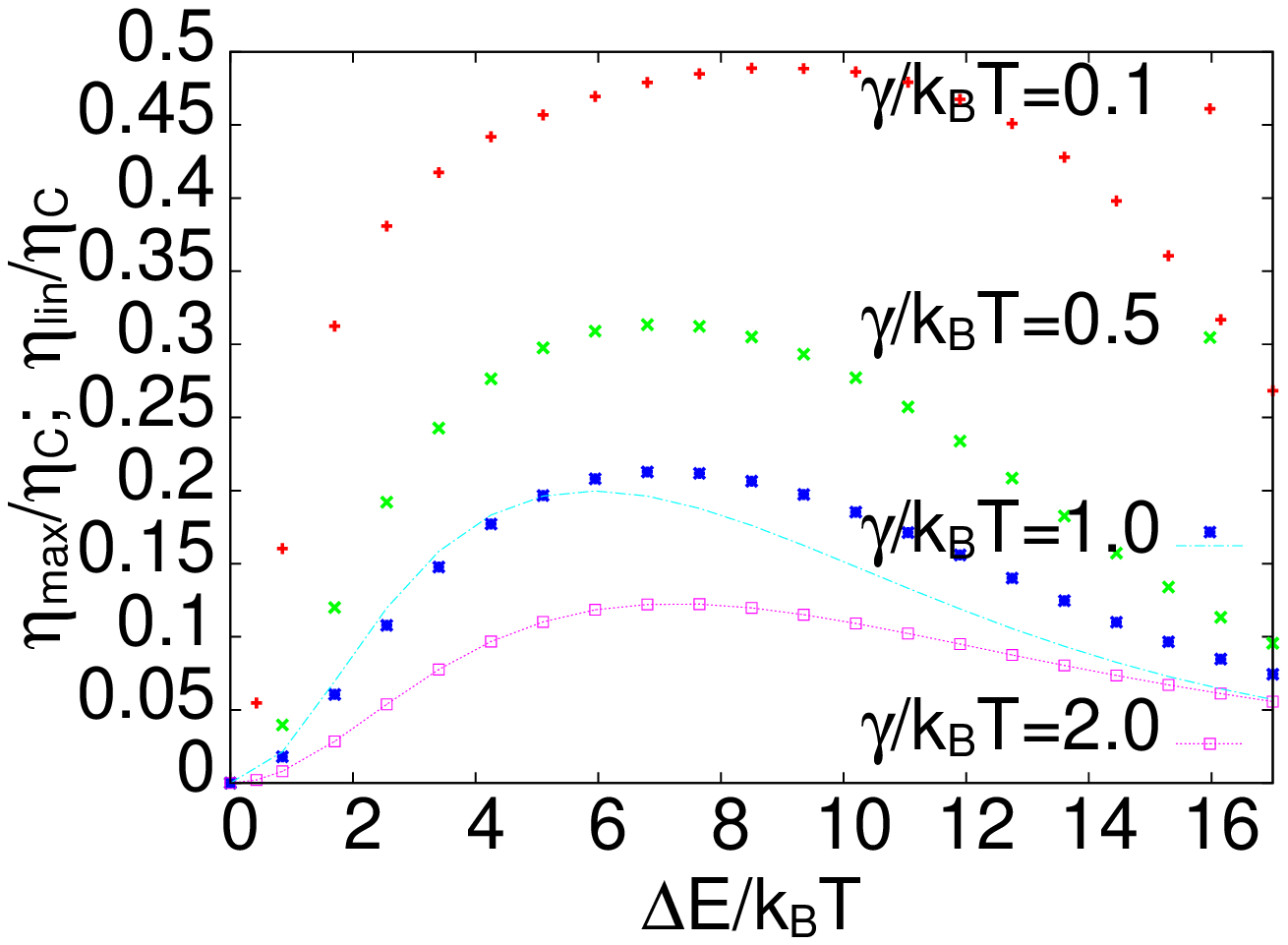}}
\caption{The dependence of the maximum power $P_{\rm max}/(k_BT)^2$
on  $\Delta E/k_BT$ for a number of $\gamma/k_BT$ values  for $\Delta T/T=0.3$ 
(left panel). Right panel: efficiency at maximum power.
The dashed line in the right panel  corresponds to $\gamma=k_BT$ and shows the 
linear efficiency $\eta_{\rm lin}/\eta_C$. }
\label{fig-maxP-eta}
\end{figure}

The most important parameters of the engine are the maximum output power $P_{\rm max}$ and the efficiency
$\eta_{\rm max}$ at maximum power. The dependence of the maximum power, scaled by $(k_BT)^2$,  on  $\Delta E/k_BT$ 
with all couplings  equal to $\gamma$  is shown in the left panel of Fig.~\ref{fig-maxP-eta}. 
Different curves correspond to different $\gamma$ in units of $k_BT$, and for each of them the power 
has been optimized with respect to the applied
voltage load. The calculations were performed  for $\Delta T/T=0.3$ with the 
non-linear effects taken into account. Interestingly, for $\Delta E/k_BT$  
up to about 12 the largest power is obtained for $\gamma=k_BT$, but for $\Delta E/k_BT>12$,
(typically) higher values of $\gamma$ lead to larger power. This non-monotonic dependence of the maximal power on 
the effective width of the resonance can be traced back to the strong $\Delta E$ dependence of the
characteristic potentials, which in turn renormalise the dots energy levels
$\varepsilon_L$ and $\varepsilon_R$ (and thus $\Delta E$). 

The right panel of Fig.~\ref{fig-maxP-eta} shows the efficiency of the engine corresponding 
to maximum power $\eta_{\rm max}/\eta_C$,
measured in units of the Carnot efficiency. The maximum value of $\eta_{\rm max}/\eta_C$ 
strongly depends on the coupling $\gamma$. 
For the optimal value of $\gamma=k_BT$ and for $\Delta T/T=0.3$, it exceeds 20\% of the 
Carnot value. The efficiency as well as the maximum power 
are increasing functions of temperature difference. The linear approximation for the 
on-dot potentials presumably precludes reliable results for $\Delta T/T>0.3$. 
For a given value of $\Delta T/T$ the maximal efficiency increases with decreasing $\gamma$, tending to the
Carnot limit when $\gamma \rightarrow 0$. At the same time the power diminishes towards zero. 
This agrees with our analysis in the linear approximation carried out in Section~\ref{subs:lin}, and 
the recent analytical treatment of the two-terminal system~\cite{yamamoto2015}. 

\begin{figure}
\centerline{\includegraphics[width=0.50\linewidth]{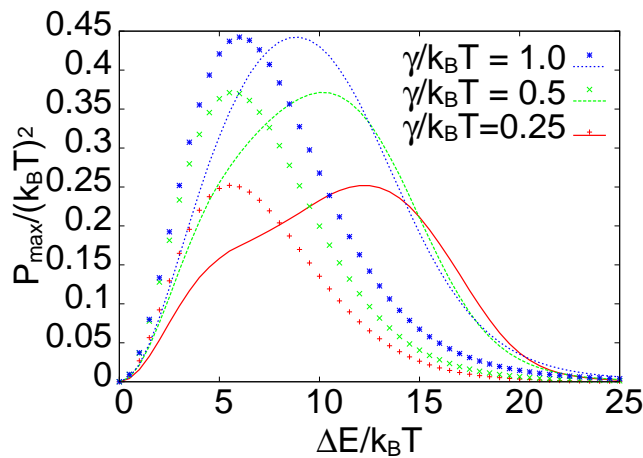}}
\caption{Comparison of the  maximum power $P_{\rm max}/(k_BT)^2$
 $vs.$ $\Delta E/k_BT$ calculated without characteristic potentials (symbols) and with them
(lines) for a number of $\gamma$ values  for $\Delta T/T=1$. }
\label{fig-compPlin-nl}
\end{figure}

In Fig.~\ref{fig-compPlin-nl} the comparison of the maximum 
power calculated with full renormalization of the on-dot energy levels (lines) 
with that obtained in the approach neglecting screening (symbols) is presented. 
Two important features are apparent. First, the non-linearities strongly affect the value 
of the dots energy difference, $\Delta E$, for which the system performs with maximum power.
While without screening effects the optimal $\Delta E$ corresponds 
to $\approx 6k_BT$, the screening shifts the optimal value to $\approx 9k_BT$. 
In order to understand this behaviour, one has to note that for a given bias $V=V_R-V_L$ and
temperature difference $\Delta T$, one may define the ``effective'' energy difference
$(\Delta E)_{\rm eff}\approx \Delta E-eu_{1,L}V-2ez_{1,L}\Delta T$; here we used
the symmetries of the $u_i$'s and $z_i$'s, as well as the smallness of the off-diagonal
potentials, cf.\ Sec.~\ref{subs:nonlin}. As a result, the upward shift of the optimal
$\Delta E$ is such that the effective energy difference roughly agrees with the optimal
value found without screening effects. In fact, the shift due to screening evaluated at
$\Delta E/k_BT=6$ is about $2k_BT$, to be compared with the difference of $\Delta E$-values
mentioned above, namely $\approx 3k_BT$, for the considered parameters ($\Delta T = T$, $\gamma = k_B T$).
The remaining difference, $\approx k_BT$, can be understood by observing that linear terms appear
in the denominators of the Green functions which are integrated over energies with the Fermi
functions in the nominator. Second, the value of the power itself and the efficiency
remain unchanged. Other differences are less important for the question of the maximum power
but should be noted. For example, one observes much stronger asymmetries of the
$P_{\rm max} (\Delta E)$ curves with the non-linear screening effects taken into account.

Comparison of the optimal power, Fig.~\ref{fig-maxP-eta}, with the power factor, 
Fig.~\ref{power-fac-gam}, shows that there exist important differences between these quantities.
In particular, the energy difference $\Delta E$ at which the power is maximal markedly differs 
from that leading to the maximum power factor: this is related to the fact that in the
full theory the voltage serves as an additional optimisation parameter.

\section{Summary and conclusions}
The  three-terminal heat nano-engine has been analysed in the linear and the non-linear approximation.
In the latter case it has been optimised with non-equilibrium  
screening effects taken into account. 
In the linear limit the Onsager symmetry relations are fulfilled due  to the floating character 
of the hot electrode. In this limit we  have calculated charge and thermal conductances, 
 the Seebeck coefficient and the thermoelectric figure of merit $ZT$. The expected efficiency calculated 
from the figure of  merit is compared with that  calculated  self-consistently in the weakly non-linear limit in
the right panel of Fig. \ref{fig-maxP-eta}. For the optimal value of the coupling ($\gamma = k_B T$) the  
linear efficiency  surprisingly well
describes the performance of the system~\cite{jordan2013} calculated exactly.
 
Our calculations show that the optimal value of the coupling is essentially unchanged by the
non-equilibrium screening effects, and equals $\gamma=k_BT$. 
These  effects do not change the maximum value of the output power and the efficiency 
obtained for the optimal value of the coupling constant and for a given value of $\Delta T/T$.
 The optimal distance between the dots energy levels, $\Delta E$, 
changes as a result of the screening effects. The change  is directly related to  modifications of the 
on-dot energy levels by the potentials $U_i$. 
The system efficiency at maximum power in units of the 
ideal Carnot efficiency exceeds 20\% for $\Delta T/T=0.3$.

In agreement with other studies of heat 
nano-engines~\cite{zebarjadi2007,muralidharan2012,meair2013,whitney2013b,szukiewicz2015}, we have found that 
the large value of the thermoelectric figure of merit $ZT$ does not necessarily
imply the usefulness of the device as an efficient energy harvester. Interestingly,
for this particular device,  we have found an overall agreement between the efficiency obtained within 
 linear approximation and in the full theory (cf.\ two sets of data for $\gamma/k_BT=1$ in 
the right panel of Fig.~\ref{fig-maxP-eta}).
Hence the three-terminal system under study markedly differs from a typical two-terminal engine.

\ack{
The work reported here has been supported by the the National Science Center grant DEC-2014/13/B/ST3/04451 (Poland),
 Deutsche Forschungsgemeinschaft (through TRR 80, Augsburg) and University of Augsburg. KIW thanks  the staff of
the Theoretical Physics II  and the  University of Augsburg for hospitality.}


\section*{References}

\end{document}